\documentclass[quaderni,english]{sifrev} 

\usepackage{graphicx}               
\usepackage{epsfig}
\usepackage[english]{babel}

\newcommand{\tvec}[1]{\mbox{\boldmath{$#1$}}}
\newcommand{\rvec}[1]{\mbox{\boldmath{${\rm #1}$}}}
\newcommand{\be}{\begin{equation}}
\newcommand{\ee}{\end{equation}}
\newcommand{\bea}{\begin{eqnarray}}
\newcommand{\eea}{\end{eqnarray}}
\newcommand{\nn}{\nonumber}

\newcommand{\oneh}{\textstyle {1\over 2}}

\begin{document}


\title{The Rise of Quantum Mechanics}

\author{Sigfrido Boffi}        
					 
\institute{Dipartimento di Fisica Nucleare e Teorica, Universit\`a degli Studi di Pavia}     

\vol{0}                                     
\issue{0}				  

\month{}                    
\year{June 2008}                               

\maketitle


\section{Introduction}
\label{sec:intro}

``The discovery and development of quantum theory in the twentieth century is an epic story and demands appropriate telling. This story cannot be told in the fullness of its glory without analyzing in some detail the multitude of problems which together came to constitute the fabric of quantum theory. Much more than the relativity theories, both special and general, which completed the edifice of classical mechanics, the quantum theory is unique in the history of science and intellectual history of man: in its conceptions it made a complete break with the past and fashioned a new worldview about the structure of matter and radiation and many of the fundamental forces of nature.'' With such emphatic words Jagdish Mehra starts a cyclopical enterprise together with Helmuth Rechenberg describing the historical development of quantum theory~\cite{mehra}. 

In fact, quantum mechanics has completely reoriented the way of looking at physical phenomena that emerged after more than three centuries of intense investigation of nature. Around the year 1900 the nowadays so-called classical physics was well organized in different sectors. Within each sector a closed and coherent system of concepts and laws was able to satisfactorily account for the corresponding phenomenology. Some remarkable syntheses, such as the  unification of electric and magnetic phenomena or the kinetic theory of matter, were suggesting that mechanics, thermodynamics, electromagnetism were only different branches of physics on the road towards a global unified description of physical phenomena. Analytical mechanics would in any case play a privileged role because the three Newton's laws were at the origin of the scientific paradigm of an objective world governed by the causality law, where the global behaviour can be lead back to the knowledge of the mutual interaction of constituents.

With the advent of quantum mechanics as a result of accounting for new facts and discoveries, this paradigm was turned over. Objectivity, determinism and locality were substituted by a picture where  the observer plays an essential role in determining the phenomenon, the description of phenomena can only be accomplished in terms of probability of occurrence, and non-locality effects have to be considered.

\begin{figure*}
\begin{center}
\includegraphics[width=11 cm]{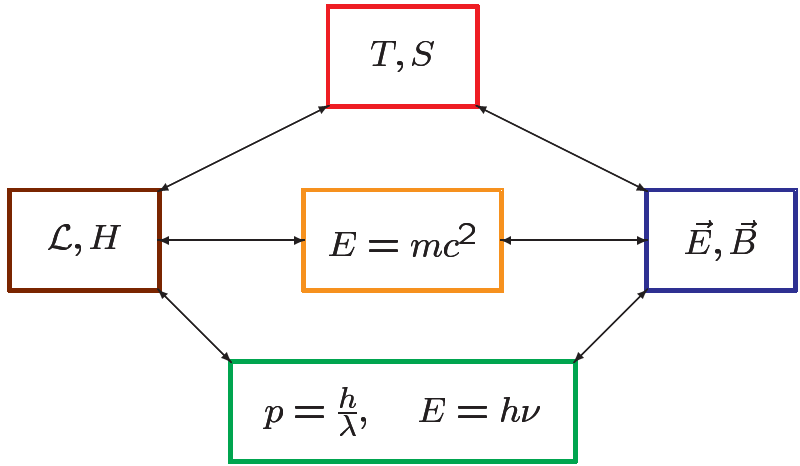}
\end{center}	  
\caption{In the first quarter of the twentieth century the crisis produced when trying to unify the different sectors of physics, such as macrophysics (described in terms of temperature $T$ and entropy $S$), mechanics (with its Lagrangian $\cal L$ and Hamiltonian $H$) and electromagnetism (with its electric  and magnetic fields $\vec{E}$ and $\vec{B}$, respectively), was overcome by introducing new concepts and a new way of thinking of reality as a consequence of the development of relativity theory (with its equivalence between energy $E$ and mass $m$ and the invariance of the light velocity $c$) and quantum mechanics (that associates, through the Planck's constant $h$, a wave with wavelength $\lambda$ and frequency $\nu$ to the motion of a particle with momentum $p$ and energy $E$, respectively).
}
\label{fig:fig1}  	        
\end{figure*}

This was achieved in the first quarter of the twentieth century, especially between June 1925 and October 1927, as a consequence of an extraordinary development of new data, ideas, formalisms, interpretations, within a polyphonic framework where very young researchers and more experienced scientists were challenging each other in a cooperative and unique effort.


\section{Crisis towards unification}
\label{sec:unity}

In analytical mechanics observers are simulated by inertial frames of reference and time is assumed to be an absolute evolution parameter. Then the objective description of phenomena means that any physical law is translated into one and the same equation when passing from one observer to another. Such a scheme suffered a big attack when physicists realized that the mechanical equations of motions are not compatible with the Maxwell's equations for the electromagnetic phenomena. The solution found in 1905 by Albert Einstein (1879--1955) with his revision of the concept of simultaneity and the space-time structure made it possible to reconcile mechanics and electromagnetism in a unified and objective picture. Thus, though revolutionary, relativity theory, even with its extension to general relativity, still obeys the principle of objectivity and lives within the paradigm of classical physics.

In contrast, in the attempt to establish a connection between the macroscopic behaviour of a complex system and the microscopic motion of its constituent particles or to account for the thermodynamic effects of radiation, one meets difficulties that are unsurmountable within the classical framework (Fig.~\ref{fig:fig1}). For example, the frequency distribution of the radiation energy density cannot be predicted invoking the classical thermodynamics of radiation. The formula proposed on heuristic arguments by Max Planck (1858--1947) in 1900 could only be explained by Einstein under the assumption that the energy of the harmonic oscillator associated to each frequency takes discrete values or, alternatively, the action corresponding to a complete oscillation is an integer multiple of an elementary value $h$, the Planck's constant. Similarly, the temperature dependence of specific heat of solids cannot be explained assuming a classical motion of atoms within the solid and violates the classical equipartition principle of energy, unless again one assumes with Einstein and Peter Debye (1884--1966) the possibility of a discrete energy spectrum for the oscillating atoms in solids.

The discrete nature of the electromagnetic field interacting with matter and Einstein's idea of a light quantum with energy $h\nu$ and momentum $h\nu/c$ were not accepted by the physics community without a long discussion.  Even after the successful test of Einstein's equation for the photoelectric effect predicting a linear relation between the maximal kinetic energy of the ejected photoelectron and the frequency $\nu$ of the incident radiation, Robert Andrews Millikan (1868--1953) remarked that ``the semi-corpuscolar theory by which Einstein arrived at this equation seems at present to be wholly untenable''~\cite{millikan}. It took other ten years to look at the light quantum as the ``photon'' responsible, e.g., of the Compton effect~\cite{lewis}.

On a different side, the discoveries of radioactivity by Wilhelm Conrad R\"ontgen (1845--1923) and of the electron in the study of cathode rays by Joseph John Thomson (1856--1940) added important insights into the constitution of matter. In atomic physics by the end of the 19th century a large amount of accumulating data on the line spectra were organized according to the combination principle emerging from the studies  of Johann Jakob Balmer (1858--1898), Johannes Robert Rydberg (1854--1919) and Walther Ritz (1878--1909). In the case of the hydrogen atom, for example, in the Balmer's formula the inverse wavelength of  every spectral line could be expressed as the difference of two terms, each of which depending on an integer number. The discrete nature of the line spectra is incompatible with the stable atom governed by the laws of classical physics, and their classification in terms of the internal atomic dynamics was a big puzzle. 

The discovery of the effect of a magnetic field on the spectral lines by Pieter Zeeman (1865--1943) and its explanation by Hendrik Antoon Lorentz (1853--1928) and Joseph Larmor (1857--1942) were a great success of the electron theory of matter. However, in some cases an anomalous line splitting was observed such as that occurring for the two sodium $D$-lines, with the $D_1$-line splitting into a quartet and the $D_2$-line into a sextet. Within the classical theory one could not explain such an anomalous Zeeman effect.

According to the model put forward in 1911 by Ernest Rutherford of Nelson (1871--1937) electrons revolving about the positively charged atomic nucleus follow a periodic motion. Quantization rules for such periodic systems were proposed in 1913 by Niels Hendrik David Bohr (1885--1962)  and implemented in 1916 by Arnold Sommerfeld (1868--1951). With such rules one defines azimuthal and radial quantum numbers describing the Kepler's orbit of the electron in a plane, and the Balmer's formula for spectral lines can be easily recovered. Also  the normal Zeeman effect could be described by Sommerfeld introducing a third quantum number, whose values determine the discrete positions of the electron orbit with respect to the external magnetic field.

The Bohr-Sommerfeld rules are derived from two postulates, i.e. the existence of stable stationary states of the atom and the definition of the emitted or absorbed radiation frequency in terms of the energy difference between initial and final stationary states. These two postulates are consequences of  the adiabatic principle and the correspondence principle. According to the adiabatic principle the quantized action remains constant during the electron motion also in the case of transitions between stationary states induced by an external perturbation. The correspondence principle implies that under suitable conditions, i.e. when the action per cycle is large compared to Planck's constant $h$, one must recover the classical limit of the radiation frequency. 

The existence of stable stationary states was confirmed in a series of experiments by James Franck (1882--1964) and Gustav Ludwig Hertz (1887--1975). The correspondence principle was a fruitful guideline in the development of the theory. Thus, scientists were confident that classical mechanics, implemented by the two Bohr's postulates and the Bohr-Sommerfeld quantization rules, could also provide the necessary foundation for atomic mechanics. 

The Bohr-Sommerfeld rules were soon applied to a variety of problems such as quantum theory of radiation, atoms with one electron and with several electrons, quantum theory of solids and gases,  atomic magnetism. They were so successful describing the constitution of atoms and the periodic table of elements that the predicted element with atomic number 72 was just discovered by Dirk Coster (1889--1950) and George de Hevesy (1985--1966) in Bohr's Institute in Copenhagen and called hafnium after the Latin name of Copenhagen ({\it Hafnia\/}), in time for Bohr to mention it in his Nobel lecture in 1922. 

However, there were also some failures, such as the calculation of the energy states of the helium atom and  similar many-electron atoms, the description of the anomalous Zeeman effect, and the difficulty to describe time dependent processes such as the interaction between radiation and matter. In the attempt to overcome the difficulty of explaining the dispersion of light by atoms Niels Bohr, Hendrik Antoon Kramers (1894--1952) and John Clarke Slater (1900--1976) assumed that a given atom in a certain stationary state communicates continually with other atoms through a mechanism which is equivalent with the classical field of radiation originating from the virtual oscillators corresponding to the various possibile transitions to other stationary states. However, the communication between atoms, i.e. the absorption and emission processes, were connected by probability laws implying that energy and momentum were conserved only on the average~\cite{BKS}. This conclusion was disproved by the result of an experiment performed by Hans Wilhelm Geiger (1882--1945) and Walter Wilhelm Georg Bothe (1891--1957)~\cite{GB} where the scattered X-ray was detected in coincidence with the recoiling electron. This result shows that energy and momentum are conserved in individual elementary processes and confirms the particle behaviour of radiation, already introduced by Einstein with the light quantum hypothesis and clearly demonstrated by Arthur Holly Compton (1892--1962)~\cite{compton}.

On the other hand, new facts and productive ideas were emerging. In 1924 in his doctoral thesis Louis-Victor de Broglie (1892--1987) suggested that waves could also be associated to particles such as electrons, a bold assumption that was most surprisingly confirmed by the findings by Clinton Joseph Davisson (1881--1958) in his studies on secondary cathode rays together with Lester Halbert Germer (1896--1971) and by George Paget Thomson (1892--1975) working on the diffraction of cathode rays by a thin film together with Alexander Reid. In G\"ottingen Walter Elsasser (1904--1987) showed that also the so-called Ramsauer effect involving the scattering of low-energy electrons by atoms of a noble gas can be interpreted on the basis of de Broglie's idea. This called for a new electromagnetism suitable to describe the wave-particle duality and its name, i.e. quantum mechanics, already appeared in the title of a paper by Max Born (1882--1970)~\cite{QM}. Thus, the systematic presentation of the results of what is now called the old quantum theory, based on the Bohr-Sommerfeld rules,  just stopped after the first paper~\cite{grundpostulate}.


\section{New formalisms}
\label{sec:forms}

In about one year from mid 1925 to mid 1926 as a result of an intensive work in G\"ottingen, Z\"urich, Cambridge and Copenhagen the situation changed dramatically. New formalisms were proposed and successfully applied to solve the problems left open by the old quantum theory. The different approaches were soon found to be equivalent, so that a complete and consistent formalism could be developed.



\subsection{Matrix mechanics}
\label{sec:matrix}

A breakthrough came with the paper conceived in June 1925 by Werner Karl Heisenberg (1901--1976) on the rocky island in the North Sea called Helgoland, where he spent a two-week vacation to recover from a hay fever attack~\cite{umdeutung}. He noticed that the formal rules in quantum theory make use of relationships between unobservable quantities, such as, e.g., the position and time of revolution of the electron. One has rather to focus on the observable quantities during emission and absorption, i.e. the radiation frequency and intensity. According to Bohr the radiation frequency $\nu$ is identified by assigning the energy of the initial and final stationary states, $W(n)$ and $W(n-\alpha)$, respectively:
$$
\nu(n,n-\alpha) = \frac{1}{h}\left[W(n) - W(n-\alpha)\right].
$$
Just in the same way one can associate a two-dimensional pattern also to the amplitude of the emitted light wave, $A(n,n-\alpha)$, with rows and columns ordered according to the different initial and final states involved in the transition. The radiation intensity is then obtained by  the squared transition amplitude. Making use of the Bohr-Sommerfeld rules and the correspondence principle Heisenberg then arrived at the following quantum condition:
\bea
h = 4\pi m\sum_{\alpha=1}^{\infty}\left[\vert A(n+\alpha,n)\vert^2 \omega(n+\alpha,n) \right.\nn\\
-\left. \vert A(n,n-\alpha)\vert^2 \omega(n,n-\alpha) \right],\nn
\eea
where $\omega(n,n')=2\pi\nu(n,n')$ and $m$ is the electron mass.

In such a reformulation scheme an essential mathematical difficulty occurred, namely the factors in product of two patterns did not commute in general. In G\"ottingen Born and Ernst Pascual Jordan (1902--1980) recognized that Heisenberg's patterns are nothing else than matrices obeying the noncommutative law of the matrix product. By sharpening the idea of the correspondence principle they adopted the classical equations of motion considering them as relations between matrices representing classical observables. That is, for a one-dimensional quantum system described by the Hamiltonian matrix $\rvec{H}=\rvec{H}(\rvec{q},\rvec{p})$ the equations of motion assume the canonical form,
\be
\dot{\rvec{q}}=\frac{\partial {\rvec{H}}}{\partial {\rvec{p}}}, \quad 
\dot{\rvec{p}}=-\frac{\partial {\rvec{H}}}{\partial {\rvec{q}}}, 
\label{eq:motion}
\ee
where the dynamical matrices $\rvec{q}$ and $\rvec{p}$ representing the system position and momentum, respectively, satisfy the quantum condition
\be
\rvec{p}\rvec{q} - \rvec{q}\rvec{p} = \frac{h}{2\pi i} \rvec{I},
\label{eq:q-cond}
\ee
with $\rvec{I}$ being the identity matrix. By repeated application of the quantum condition (\ref{eq:q-cond}) the equations of motion~(\ref{eq:motion}) could equivalently be rewritten as
\be
\dot{\rvec{q}}=\frac{2\pi i}{h}(\rvec{H}\rvec{q}-\rvec{q}\rvec{H}), \quad 
\dot{\rvec{p}}=\frac{2\pi i}{h}(\rvec{H}\rvec{p}-\rvec{p}\rvec{H}). 
\label{eq:move}
\ee

Together with Heisenberg they extended the scheme to include systems having arbitrarily many degrees of freedom and developed a quantum-mechanical perturbation theory. Their paper was soon cited as the {\it Dreim\"annerarbeit\/} (three-men's paper), and the theory they developed was called matrix mechanics~\cite{drei}. In particular, in complete analogy with the classical Hamilton-Jacobi equation they found that the energy levels could be derived by transforming the Hamiltonian matrix $\rvec{H}$ to its diagonal form $\rvec{W}$ by using  a (unitary) transformation matrix $\rvec{S}$:
\be
\rvec{H}(\rvec{q},\rvec{p})= \rvec {S} \rvec{H}(\rvec{q}_0,\rvec{p}_0)\rvec{S}^{-1} = \rvec{W}.
\label{eq:diag}
\ee
When the transformation with $\rvec{S}$ is applied to the matrices $\rvec{q}$ and $\rvec{p}$, the quantum condition (\ref{eq:q-cond}) is left invariant. This justifies the name of canonical transformation given to it by Born, Heisenberg and Jordan. 

Canonical transformation and diagonalization of the Hamiltonian matrix $\rvec{H}(\rvec{q},\rvec{p})$ are the central ideas of the three men's paper. They help to find all the conserved quantities for a given quantum system as those represented by matrices that have a vanishing commutator with $\rvec{H}(\rvec{q},\rvec{p})$. This is in turn in close analogy with classical mechanics where an integrable system with $f$ degrees of freedom has exactly $f-1$ independent constants of motion besides the Hamiltonian.

The first application to a physical problem was successfully accomplished by Wolfgang Pauli (1900--1958) who was able to derive the correct spectrum of the hydrogen atom after a brilliant and laborious calculation~\cite{hydrogen}, where he showed that also in the quantum-mechanical case both angular momentum and the Runge-Lenz vector are constants of motion.



\subsection{Wave mechanics}
\label{sec:wave}

On a different side, the new electromagnetism wished by De Broglie was formulated by Erwin Schr\"odinger (1887--1961) in four papers produced between January and June 1926 where quantization was developed as an eigenvalue problem~\cite{erwin}. Schr\"odinger took advantage of the analogy between the wave propagation and the motion of a particle already explored by de Broglie in his thesis and even earlier by William Rowan Hamilton (1805-1865) in 1824. Wave propagation of light can be visualized according to Christian Huyghens (1629--1695) by looking at the motion of the wave front, i.e. the surface with constant wave phase perpendicular to the wave vector; alternatively, according to Pierre de Fermat (1601--1665) one can describe the propagation in terms of a light ray always tangent to the wave vector. The bending of the ray and the distortion of the phase wave are both due to local variations of the refractive index. In quite a similar way the motion of a particle in terms of its trajectory, always tangent to the particle momentum, can also be visualized in terms of an action wave, always perpendicular to the particle momentum. Modulations of the potential affect the momentum just like the refractive index modifies the wavelength. 

These similarities were already summarized in the Einstein-Planck formula $E=h\nu$ and the de Broglie's hypothesis $p=h/\lambda$, where particle quantities, such as the energy $E$ and momentum $p$, were connected through Planck's constant with wave quantities, such as the frequency $\nu$ and wavelength $\lambda$. 

Assuming relativistic kinematics, as also de Broglie did, Schr\"odinger first derived an equation which is nowadays known as the Klein-Gordon equation, but he gave it up because it did not yield the right fine-structure of the hydrogen atom. In fact, the equation entitled after Oskar Benjamin Klein (1894--1977) and Walter Gordon (1893--c.1940) has many fathers~\cite{kragh} and was recovered a few years later by Pauli and Victor Frederick Weisskopf (1908--2002) who gave it the correct interpretation within a newly developing quantum field theory~\cite{vicki}.

Confining himself to nonrelativistic kinematics, in the first paper of the series~\cite{erwin} Schr\"odinger considered an electron bound in the hydrogen atom. In this case the Bohr-Sommerfeld quantization rules imply stationary waves with wavelength tuned to the orbit length. Then the Hamilton-Jacobi equation of analytical mechanics becomes
\be
\nabla^2\psi + \frac{2m}{K^2}\left(E + \frac{e^2}{r}\right)\psi = 0,
\ee 
where $K$ has the dimension of an action (nowadays $K\equiv\hbar=h/2\pi$) and $\psi$ describes the wave to be associated with the electron motion with energy $E$. Therefore the Hamilton-Jacobi equation has turned into an eigenvalue equation for the electron Hamiltonian. Its solutions correspond to the stationary states of Bohr's atomic theory and provide the discrete spectrum of the bound electron, in good agreement with data. This was a much more direct approach to the problem than the quite complicated calculation of Pauli with matrix mechanics. In modern textbooks the nonrelativistic quantum hydrogen problem is solved according to the procedure followed by Schr\"odinger.

The analogy between particle motion and geometrical optics was further examined in the second communication, where the eigenvalue equation was derived from a variational principle and applied to other soluble cases such as the Planck's oscillators and the rigid rotator. In the third paper the method could also be applied in perturbation theory to cases where exact analytical solutions are impossible, such as the Stark effect. Only in the fourth communication the process of building the new wave mechanics was accomplished with the introduction of the time-dependent Schr\"odinger equation:
\be
\nabla^2\psi - \frac{8\pi ^2}{h^2} V\psi + \frac{4\pi i}{h}{\partial\psi\over\partial t} = 0
\label{eq:erwin}
\ee
(with $m=1$). This was obtained by replacing the energy $E$ with the operator $i(h/2\pi)\partial/\partial t$ acting on the wave function $\psi$ and assuming that the potential energy $V$ works on $\psi$ as a multiplicative operator. Similarly, the kinetic energy is responsible for the Laplace operator $\nabla^2$ coming from the replacement $p\to-i(h/2\pi)\nabla$. Thus, as we do it today, eq.~(\ref{eq:erwin}) can equivalently be written as
\be
i\hbar\frac{\partial\psi}{\partial t} = H\psi,
\label{eq:schrodinger}
\ee
where $H$ is the Hamilton operator. For an isolated system its most general solution is given by a linear superposition of particular solutions,
\be
\psi = \sum_n c_n u_n \,e^{-iE_n t/\hbar}.
\label{eq:general}
\ee
where $u_n$ and $E_n$ are the eigenfunctions and eigenvalues, respectively, of $H$, i.e.
\be
H u_n = E_n u_n.
\label{eq:eigen}
\ee
Mathematically, Eq.~(\ref{eq:general}) is dictated by linearity of Schr\"odinger's equation. Physically, it  is a wave function reflecting the linear superposition principle typical of a wave behaviour.



\subsection{Equivalence between matrix and wave mechanics}
\label{sec:equivalence}

During a stay at MIT in the winter semester 1925--1926 to take advantage of the collaboration with Norbert Wiener (1894--1964), the future father of cybernetics, Born realized that also the Hermitian matrices of matrix mechanics could be regarded as operators acting on vectors in a multidimensional space. Assuming the Hamiltonian to be an operator function of the dynamical variables, having the same functional dependence on the operators $\rvec{p}$ and $\rvec{q}$ as the classical Hamiltonian has on its dynamical variables, one could reformulate the laws of matrix mechanics for any system~\cite{wiener}.

In March 1926, just after his second communication and before his third one, Schr\"odinger was able to show the link between wave mechanics and matrix mechanics~\cite{equivale} claiming that ``from the  formal mathematical standpoint one may even say that the two theories are {\it identical\/}''. In fact the matrix elements of the Hermitian matrices representing operators in matrix mechanics are just the same elements obtained using the wave functions and the operators in wave mechanics.

Also Pauli, in a letter to Jordan dated April 12, established the connection between wave and matrix mechanics showing that ``the energy values resulting from Schr\"odinger's approach are always the same as those of the G\"ottingen Mechanics, and that from Schr\"odinger's function $\psi$, which describes the eigenvibrations, one can in a quite simple and general way construct matrices satisfying the equations of the G\"ottingen Mechanics''. He never published the content of his letter that was discussed in public only many years later~\cite{waerden}. 

It may be of some interest to recall that already at the end of December 1925  also Cornelius Lanczos (1893--1974) arrived at an integral equation equivalent to Schr\"odinger's equation starting from the matrix mechanics of Heisenberg, Born and Jordan and applying Hamilton's variational principle~\cite{lanczos}. His paper, however, was not appreciated by Schr\"odinger and Pauli and remained isolated in Lanczos' production.

For completeness, one should mention that also Carl Henry Eckart (1902--1973), after attending Born's lectures during his tour in U.S.A. in winter 1926, was able to prove the equivalence between matrix and wave mechanics~\cite{eckart} just before publication of Schr\"odinger's paper.



\subsection{Dirac's $q$-numbers}
\label{sec:dirac}

After reading the first article of Heisenberg~\cite{umdeutung} Paul Adrien Maurice Dirac (1902--1984) realized that the new theory was suggesting ``that it is not the equations of classical mechanics that are in any way at fault, but that the mathematical operations by which physical results are deduced from them require modification. {\it All\/} the information supplied by the classical theory can thus be made use of in the new theory''~\cite{fundamental}. This statement followed from the fact that Dirac recognized the commutation relations between quantities representing observables to have similar properties as the Poisson's brackets of classical mechanics. To distinguish classical variables from the quantum noncommuting objects ``we shall call the quantum variables $q$-numbers and the numbers of classical mathematics which satisfy the commutative law $c$-numbers, while the word number will be used to denote either a $q$-number or a $c$-number''~\cite{prelim}. Thus, replacing the Poisson's bracket $\{A,B\}$ of two classical observables by the commutator $[A,B]=AB-BA$ of the two corresponding $q$-numbers,
\be
\{A,B\}\to -\frac{i}{\hbar}[A,B],
\ee
Dirac could extend the Hamilton formalism to quantum equations of motions~\cite{fundamental}. 

``The new quantum mechanics consists of a scheme of equations which are very closely analogous to the equations of classical mechanics, with the fundamental difference that the dynamical variables do not obey the commutative law of multiplication, but satisfy instead the well-known quantum conditions. It follows that one cannot suppose the dynamical variables to be ordinary numbers ($c$-numbers), but may call them numbers of a special type ($q$-numbers). The theory shows that these $q$-numbers can in general be represented by matrices whose elements are $c$-numbers (functions of a time parameter)''~\cite{dirac-interpret}.

In a series of eleven papers published in twenty months, without particularly new results,  Dirac was able to give an extraordinary new perspective in the formal and conceptual development of the new quantum mechanics. 

In particular, for a multiply periodic system action and angle variables, $J_k$ and $w_k$ respectively, could be introduced satisfying equations of motion formally identical to the classical equations, i.e.
$$
\dot J_k = [J_k,H] = 0, \quad \dot w_k = [w_k,H] = \frac{\partial H}{\partial J_k}.
$$
An application to the hydrogen atom~\cite{prelim} immediately followed that given by Pauli with the matrix mechanics~\cite{hydrogen} confirming his results. Then Dirac extended the action-angle scheme to investigate many-electron atoms; he recovered the Land\'e formula for the anomalous Zeeman effect  as well as the relative intensities of the spectral lines in a multiplet and their components in the presence of a weak magnetic  field. The only new result was an application to the Compton effect, where his ``theory gives the correct law of variation of intensity with angle, and suggests that in absolute magnitude Compton's values are 25 per cent too small''~\cite{dirac-compton}. Indeed, a few months later in a letter to Dirac~\cite{letter-a}, Compton announced that new observations were then in quite good agreement with theory!

In Dirac's opinion a good notation and a clear nomenclature are essential tools. Therefore he invented new terms and symbols, some of them still in common use today, such as `commutator', `$q$-numbers', `eigenfunction', the `$\delta$-function'. The famous bra  and ket notation appeared only much later~\cite{braket} and was used in the third edition of the celebrated book~\cite{diracbook}, originally published in 1930 and practically unmodified in ten over twelve chapters up to the last (fourth) edition in 1958, remaining a fundamental reference for any beginner also today.

Dirac's algebraic approach in terms of $q$-numbers may be considered a generalization of the matrix mechanics suitable for both periodic and aperiodic motions. It also provides a unified formalism for the new quantum theory, because the $q$-numbers can be related to the operators of the Born-Wiener approach and Schr\"odinger's wave mechanics by building their matrix representation~\cite{diracstat}.

Dirac's quantum algebra makes use of what is now called the abstract Hilbert space. This is a linear manifold of vectors (i.e. closed under vector addition and multiplication by scalars, and with a strictly positive inner product over the field of complex numbers) which is complete with respect to the metric generated by the inner product and separable. Its elements are legitimate objects to represent physical states, and observables correspond  to suitable linear (self-adjoint) operators acting on such elements. A basis in the Hilbert space is given by the complete set of eigenvectors of one of these self-adjoint operators, so that the vector representing the state of the system can be written as a linear superposition of the basis elements. This fact reflects the linear superposition principle which has a central role in quantum mechanics to describe the wave-particle duality.

This scheme was brought to a precise formulation by the G\"ottingen mathematical school flourished around David Hilbert (1862--1943),  in particular by Johannes (John) von Neumann (1903--1957)~\cite{neumann}.



\section{The wave function and transformation theory} 
\label{sec:transform}

The existence of a continuity equation for $\rho=\vert\psi\vert^2$ associated with his wave equation, in quite analogy with hydrodynamics, induced Schr\"odinger in his fourth communication to assume $\rho$ to describe the matter distribution of the particle (an electron of total charge $e$) and $e\rho$ its electric charge distribution. This idea was strengthened by finding that a suitable wave packet built as a superposition of harmonic oscillator eigenfunctions could remain concentrated during its motion through space with a back-and-forth time behaviour just as in classical case~\cite{stetige}~\footnote{%
Schr\"odinger's wave packet is the minimum uncertainty (coherent) state introduced by Roy Jay Glauber (b. 1925) many years later to describe the laser radiation field~\cite{glauber}.
}. %
This interpretation was immediately rejected by Erwin Madelung (1881-1972)~\cite{madelung} because in the Hamiltonian driving the Schr\"odinger's equation there is no mutual interaction between different parts of the particles distributed over the whole space. For both Schr\"odinger and Madelung the quantity $\rho$ is in any case a weight function necessary to calculate average values just as we do it today.

In contrast, de Broglie gave a realistic interpretation assuming that the wave function $\psi$ is a physically real wave, a pilot wave driving the particle during its motion. Its velocity is derived by the guidance formula, $\tvec{v}=\tvec{\nabla}S/m$, and is always perpendicular to a surface of constant action $S$ determined by $\psi$~\cite{broglie}. Through the definition of a quantum potential to be added to the usual potential in the classical equation of motion, de Broglie intended to recover the deterministic behaviour of classical mechanics in terms of variables that remain hidden in the theory. Such an interpretation was so strongly confuted by Pauli at the Fifth Solvay Conference in Bruxelles in October 1927 that de Broglie abandoned it for many years and resurrected it only when David Bohm (1917--1992) proposed his approach to hidden variables in 1952~\cite{bohm}.

``As a matter of fact, the new mechanics does not answer, as the old one, to the question {\it how does the particle move\/}, but rather to the question {\it how probable is that a particle moves in a given way\/}''~\cite{born}. This revolutionary statement follows Born's discovery that in scattering processes in particle collisions the wave function $\psi$ plays the same role as the electric field $\tvec{E}$ in light diffraction. What matters to describe the angular distribution of particles and light is neither $\psi$ nor $\tvec{E}$, but rather their squared modulus. The arrival of a particle (or a photon) at some point on a screen is not predictable, only its probability can be calculated with $\vert\psi\vert^2$. Thus the wave function acquires a statistical interpretation and is well recognized to have only a pure auxiliary role in the calculation of observable quantities~\cite{stoss}.

On the other hand, within the formalism the wave function provides a complete and exhaustive description of the system under consideration, as shown by Dirac in his transformation theory~\cite{dirac-interpret}. Originally, this approach is an elegant solution to the general problem of solving the quantum equations of motion either in matrix or in wave mechanics. Similar ideas were proposed independently at the same time by Jordan~\cite{jordan-a,jordan-b} and Fritz Wolfgang London (1900--1954)~\cite{london}. 

Equations of motion in quantum mechanics, in the form of either Eqs.~(\ref{eq:move}) or Eq.~(\ref{eq:schrodinger}), involve the Hamiltonian, i.e. a Hermitian (or, better, self-adjoint) operator. In both cases one has to construct a matrix representation of the Hamiltonian and apply a suitable transformation to bring it to a diagonal form by solving either Eq.~(\ref{eq:diag}) or Eq.~(\ref{eq:eigen}). Already in the three-men's paper such a transformation was found to be unitary. Dirac extended the procedure by showing that in a scheme of matrices representing the dynamical variables unitary transformations preserve all algebraic relations such as the commutation relations, the equations of motion and the expectation values. Therefore, the transformed set of matrices is just equivalent to the original one, so that it is a free choice to adopt a scheme where, e.g., the position (the (q) scheme) or the energy (the Hamiltonian) are diagonal. ``The eigenfunctions of Schr\"odinger's wave equation are just the transformation functions \dots that enable one to transform from the (q) scheme of matrix representation to a scheme in which the Hamiltonian is a diagonal matrix''~\cite{dirac-interpret}. 

Commuting matrices can be put in diagonal form in the same scheme. In the scheme where the Hamiltonian is diagonal only a few set of matrices can also be brought to diagonal form: the corresponding dynamical variables are the only possible constants of motion. At any time the state of the physical system with a precise value of energy is then fully characterized by assigning also the values of such constants. For other variables, not commuting with the Hamiltonian, one can only calculate their average value in that state: ``this information appears to be all that one can hope to get''~\cite{dirac-interpret}. 


\section{A new degree of freedom}
\label{sec:newdof}

During the same couple of years when the basic formalism of quantum theory was developing, a new degree of freedom entered the scene of atomic physics. It immediately appeared to be the last remaining building element in the puzzle to fit the data.


\subsection{Spin}
\label{sec:spin}

In order to account for the periodic system of elements, something was still missing. At the time, electrons in an atom were divided in two groups, one of which living in the passive atomic core and the other defining the position of that atom in a series of the periodic system. Such series electrons were considered responsible for all atomic properties including magneto-mechanical effects and radiative processes. Their distribution among atomic levels could explain, e.g., the ground state structure of noble gases~\cite{stoner}. However, in the attempt to explain the multiplet structure and the selection rules of the anomalous Zeeman effect by refining Sommerfeld's approach, Alfred Land\'e (1888--1975) and Werner Heisenberg already in 1921-1922 realized that half-integral quantum numbers had to be used for the series electrons. Also Pauli during his stay in Copenhagen by Bohr in 1923 convinced himself that for elements that follow each other in the periodic table the values of the magnetic quantum number are alternatively half-integral and integral. He then speculated that this was caused ``by a peculiar, classically not describable kind of duplicity of quantum-theoretical properties of the series electron''~\cite{pre-spin} demanding the introduction of a fourth quantum number in the classification of electron orbits. In the following discussion of the problem of equivalent electrons, i.e. electrons having the same binding energy (or the same principal quantum number), he arrived at the conclusion that ``there can never exist two or more equivalence electrons in the atom for which \dots the values of all [four] quantum numbers \dots coincide''. The Pauli exclusion principle was immediately accepted by the physicists working in the field and became clearer when George Eugene Uhlenbeck (1900--1988) and Samuel Abraham Goudsmit (1902--1978) formulated the spin hypothesis associating Pauli's fourth quantum number with ``an intrinsic rotation of the electron''~\cite{spinhyp}. 

According to Uhlenbeck and Goudsmit an intrinsic magnetic moment has to be associated with each electron,
$$
\tvec{\mu}_S = - \frac{e}{mc} \tvec{s},
$$
where $\tvec{s}$ is the angular momentum of the intrinsic rotation, the spin vector of the electron, and the gyromagnetic ratio $\tvec{\mu}_S / \tvec{s}$ is assumed to be twice as large as the one for the orbital motion. Taking into account this new degree of freedom Heisenberg and Jordan were able to finally describe the anomalous Zeeman effect correctly~\cite{anomalous} including the spin-orbit interaction due to the internal magnetic field felt by the electron moving in the Coulomb field of the nucleus, as explained by Llewellyn Hilleth Thomas (1903--1992)~\cite{thomas}. In addition, the fine structure of the one-electron atom could be reproduced considering the effect of the spin-orbit interaction together with the relativistic correction of the kinetic energy to order $p^4$. The result for an atom with atomic number $Z$, 
$$
\Delta E = \frac{2R^2h^2Z^4}{mc^2n^3} \left [ \frac{3}{4n} - \frac{1}{j+\oneh} \right ]
$$
(with the Rydberg's constant $R= 2\pi^2e^4m/h^3$), gives the shift of the energy levels characterized by the principal quantum number $n$ and total spin $j$. Still the first excited state of the hydrogen atom remains degenerate: today we know that the splitting between the $n^{2s+1}L_j = 2^2S_{1/2}$ and $2^2P_{1/2}$ states, the so-called Lamb shift~\cite{lamb}, can only be explained within a completely relativistic quantum field theory approach.

The spin was soon incorporated in the emerging formalism of wave mechanics by Pauli, who represented the electron spin as an operator with the same formal properties of angular momentum. Introducing the famous Pauli matrices, he derived the Schr\"odinger's equation for a particle interacting with an external magnetic field~\cite{spin}. However, spin has a relativistic origin, as realized by Dirac in connection with his equation for the relativistic electron~\cite{electron}, and Pauli's equation can be derived as the nonrelativistic limit of Dirac's equation.



\subsection{Spin statistics}
\label{sec:statistics}

A derivation of Planck's formula was given by Satyendra Nath Bose (1894--1974) in 1924 using only the corpuscolar picture without any reference to wave-theoretical concepts~\cite{bose}. The paper, sent to Einstein with the request to sponsor its publication, was enthusiastically translated into German by Einstein himself  and submitted to {\it Zeitschrift f\"ur Physik\/}. It also inspired Einstein to give an analogous application to the theory of the so-called degeneration of ideal gases~\cite{einsteinstat}, now known to describe the thermodynamical properties of a system of particles with symmetrical wave functions according to what is called the Bose-Einstein statistics.

The spin hypothesis, together with Pauli exclusion principle, was also of great help in solving the problem of the two-electron atom and the stability of ortohelium. This is an excited state of the helium atom in a triplet configuration where the spin of the two electrons are aligned in a symmetrical spin configuration with total spin $S=1$. Without considering spin orthohelium would be degenerate with parahelium, where the spins of two electrons are antiparallel and live in a singlet (antisymmetric) state with $S=0$. Degeneration is removed and orthohelium becomes more bound when antisymmetry of the total wave function under interchange of the two electrons is taken into account~\cite{helium}, in agreement with the rule formulated by Friedrich Hund (1896--1997) concerning the lower energy of states with higher spin value~\cite{hund}. In addition, for symmetry reasons ortohelium can hardly decay to the ground (singlet) state, thus explaining its stability.

The Pauli exclusion principle was also applied by Enrico Fermi (1901--1954) to molecules in a quantum gas~\cite{fermi}. A few months later, Dirac found that symmetrical wave functions of a system of identical particles lead to the Bose-Einstein statistical mechanics, and antisymmetrical wave functions satisy Pauli principle~\cite{diracstat}. Therefore, spin-$\oneh$ particles, like electrons in an atom, are said to obey Fermi-Dirac statistics. 

Incidentally, the determinantal form of the antisymmetric wave function of several independent electrons, now known as the Slater determinant~\cite{slater}, was used for the first time just by Dirac in Ref.~\cite{diracstat}. It was also adopted by Vladimir Alexsandrovich Fock (1898--1974)~\cite{fock} to refine the mean field approach to the many-electron atom proposed by Douglas Raynes Hartree (1897--1958)~\cite{hartree}. The Hartree-Fock method opened the road towards understanding atomic structure from the basic quantum theoretical principles~\cite{condonshortley}.

The spin hypothesis together with Fermi-Dirac statistics found one of its first applications when Pauli explained the paramagnetism of the electrons in a metal~\cite{pauli}. The general form of the connection between spin and statistics was proven by Pauli some years later within the frame of quantum field theory~\cite{pauli-stat}. It states that particles with integer or half-integer spin must be quantized according to Bose-Einstein or Fermi-Dirac statistics, respectively. 


\section{Indeterminacy principle}
\label{sec:indet}

By the end of 1926 and beginning of 1927 a critical analysis of the formalism lead to the discovery of an in principle difficulty to determine the values of all the independent dynamical variables required by the system degrees of freedom to specify its state.

In classical mechanics it is assumed that the position $q_r$ of the $r$-th particle in a system can be determined together with its momentum $p_r$ at a specific instant of time. Such a knowledge allows one to follow the particle motion by looking at its trajectory trough space according to the classical equations of motion. In contrast, by the end of 1926 Dirac found that, as a consequence of the commutative law of multiplication existing among $q$-numbers, in the quantum theory it is impossible to specify the value of any ``constant of integration'' by numerical values of the initial coordinates and momenta $q_{r0}$ and $p_{r0}$. ``One cannot answer any question on the quantum theory which refers to numerical values for both $q_{r0}$ and $p_{r0}$. One would expect, however, to be able to answer questions in which only the $q_{r0}$ or only the $p_{r0}$ are given numerical values''~\cite{dirac-interpret}~\footnote{%
A similar conclusion was reached also by Jordan: ``with a given value of $q$ all possible values of $p$ are equally probable''~\cite{jordan-b}.}. %

When writing this paper Dirac was in Copenhagen where he presented his ideas in a seminar. Three months later, Heisenberg, who attended the seminar, delivered his famous paper on the intuitive ({\it anschaulich\/}) content of kinematics and mechanics~\footnote{%
According to Immanuel Kant (1724--1804) in his {\it Kritik der reinen Vernunft\/}, the German word {\it Anschauung\/} is the intuition, or knowledge, that results from the immediate apprehension of an independently real object. Ethimologically, the word corresponds to English ``to look at'', an active process to grasp the meaning of some observed fact. It corresponds to the Latin ``intuere'', therefore {\it anschaulich\/} is better translated as ``intuitive''. In contrast, the often used word ``evident'', like the Latin ``e-videre'', is more suited to describe a passive role of the observer who acquires his knowledge as emerging to his consciousness from the phenomenon itself. Heisenberg, who held Kant in high esteem, was particularly sensitive to {\it Anschaulichkeit\/}, a necessary property for him to describe the physical world. However, for him {\it Anschaulichkeit\/} could not refer to the lost classical and causal space-time description. ``We believe that we intuitively understand the physical theory when we can think qualitatively about individual experimental consequences and at the same time we know that  application of the theory never contains internal contradictions''~\cite{indetermina}.
\newline
As a comment, we have to admit that the intuitive description in quantum mechanical terms is not always evident. Schr\"odinger, for example, felt {\it abgeschreckt\/} (discouraged) by the abstract approach of matrix mechanics~\cite{equivale}, whereas Heisenberg found Schr\"odinger's approach {\it abscheulich\/} (disgusting) and his claim about its {\it Anschaulichkeit\/} a {\it mist\/} (rubbish)~\cite{letter-b}.
}. %
Inquiring what information can be derived  from the Dirac-Jordan's transformation theory he discovered that canonically conjugate variables such as the position and momentum of a particle cannot be exactly determined simultaneously. There is rather an indeterminacy relation between the precision $\Delta q$ in position and the precision $\Delta p$ in momentum involving the Planck's constant~\cite{indetermina}, i.e.
\be
\Delta p\ \Delta q \ge \frac{h}{4\pi}.
\label{eq:indet}
\ee
Relation (\ref{eq:indet}) found its counterpart in a careful scrutiny of the measurement process of position and momentum of an electron that unavoidably has to involve physical phenomena such as Compton effect and wave diffraction. Therefore, relation (\ref{eq:indet}) reflects an indeterminacy principle characterizing physics. ``The more accurately the position is determined, the less accurately the momentum is known and conversely''~\cite{indetermina}~\footnote{%
Throughout the paper Heisenberg used the word {\it Ungenauigkeit\/} (imprecision) rather than {\it Unbestimmtheit\/} (indeterminacy) or {\it Unsicherheit\/} (uncertainty) that were later on also used by him. As a matter of principle, relation (\ref{eq:indet}) states that it is impossible to simultaneously and precisely {\it determine\/} position and momentum. Therefore, the word `indeterminacy' should be preferred. The word `uncertainty', in current use in English written textbooks, reminds us of our feeling rather than of the result of an observation.}. %
The classical concept of trajectory with its sharp definition of position and momentum at any time becomes meaningless, and ``in the strong formulation of the causal law `If we know exactly the present, we can predict the future' it is not the conclusion but rather the premise which is false. We cannot know, as a matter of principle,  the present in all its details''. As a consequence, ``because all experiments are subject to the laws of quantum mechanics and therefore to equation [(\ref{eq:indet})], the invalidity of the causal law is definitely established in quantum mechanics''.

For a quantum particle the minimum indeterminacy, corresponding to equal sign in (\ref{eq:indet}), is gained when it is described by a wave packet with Gaussian form~\cite{kennard}, where  $\Delta q$ ($\Delta p$) represents the width of the packet in configuration (momentum) space. 

In reply to an objection raised by Edward Uhler Condon (1902--1974) suggesting that in some cases conjugate variables could be determined simultaneously~\cite{condon}, relation (\ref{eq:indet}) was shown by Howard Percy Robertson (1903--1961) to be a particular case of a more general relation where the state of the system also plays a role~\cite{robertson} (see also~\cite{prussian}), i.e.
\be
\Delta A \ \Delta B \ge \oneh\vert\langle [A,B]\rangle \vert ,
\ee
where $\Delta A$ is the variance of the distribution of the possible values of the dynamical variable $A$ for the state of the system, and the nonvanishing average value of the commutator $[A,B]$ indicates that the two observables associated with the self-adjoint operators $A$ and $B$ are not compatible, i.e. cannot simultaneously be determined with any accuracy. 

According to the picture emerging from the Dirac's and Heisenberg's papers, before measurement the system could be in any eigenstate of $A$, in fact in a linear superposition of eigenstates. After measurement, among all possible outcomes one value of $A$ is selected, say $\alpha$, and the system is `projected' onto the eigenstate belonging to $\alpha$.  Another measurement of the variable $B$, immediately performed after the first one, would project the new state onto an eigenstate of $B$. When $A$ and $B$ do not commute, the new state is different and the information on $A$ is lost. Contrary to classical physics the second measurement does not enrich our information on the system.

The irreversible projection pheno\-menon, not predictable by Schr\"odinger's equation, is known as the collapse, or reduction, of the wave packet. It was promoted to an explicit postulate of the formalism by von Neumann~\cite{neumann}.

Analyzing a Stern-Gerlach experiment, Heisenberg also showed that the precise determination of energy is higher, the larger is the time spent by the atom crossing the deviating magnetic field, i.e.
\be
\Delta E\ \Delta t \sim h,
\ee
so that there is also an indeterminacy principle for the conjugate variables energy and time, although time is keeping in quantum mechanics the same parametric role as in classical physics.


\section{The physical interpretation}
\label{sec:interpret}

The mathematical framework of quantum theory was basically fixed in 1932~\cite{neumann}. However, since its first appearance and already during its development the new formalism was extensively and successfully applied to atomic physics. With the discovery of the indeterminacy principle, imposing to definitely abandon the space-time causal description, it soon became clear that also epistemological and ontological problems had to be addressed. 

In quantum theory, as in any physical theory, one may distinguish three components. There is the formalism with its primitive notions and axioms from which one logically derives a set of formulae. In order to be physically meaningful these formulae have to be correlated with observable phenomena by a set of rules of correspondence. This correlation is ultimately intended to establish physical laws referring to both describing what is observed and predicting new data. The formalism and the correspondence rules make quantum theory first of all a procedure: it is ``a procedure by which scientists predict probabilities that measurements of specified kinds will yield results of specified kinds in situations of specified kinds'''~\cite{stapp}. Accordingly, the quantum theoretical formalism is to be interpreted pragmatically. ``The task of science is both to extend the range of our experience and to reduce it to order''~\cite{order}. Specifically, quantum theory does not describe a system in itself, but only deals with the results of actual observations on it. Thus, particular attention has to be paid to the measurement process because deciding the kind of measurement already means to emphasize one of the complementary aspects of the quantum system. In order to determine the position and momentum of a particle, mutually exclusive experimental arrangements must be made use of. While the measurement device and its operation are specified by classical physics concepts, the mathematical formalism of quantum mechanics offers rules to calculate expectations about observations.

The third component is the interpretation of the theory. This involves philosophical issues, such as the questions about the physical reality and its description within the theory. In the history of physics for the first time interpretation has acquired particular importance when dealing with quantum mechanics. The reason is that with it the classical picture of a real world to be described objectively has been entirely turned over by the linear superposition principle and the indeterminacy principle. Morevover, as Heisenberg and Bohr were claiming, quantum theory provides us for a complete scientific account of atomic phenomena. ``The essentially new feature in the analysis of quantum phenomena is \dots the introduction of a fundamental distinction between the measuring apparatus and the objects under investigation \dots While within the scope of classical physics the interaction between the object and apparatus can be neglected or, if necessary, compensated for, in quantum physics this interaction thus forms an inseparable part of the phenomenon''~\cite{fenomeno}. Therefore, the phenomenon, i.e. what appears to us, is not merely a manifestation to our senses (even powered by sophisticated instruments) of an objective reality, which is absolute and independent of the observer, with its deterministic laws. It is rather the encounter between the observed and the observer: it is the result of an autonomous decision of the scientist to look at one of the complementary aspects with the kind of apparatus he has chosen. Consequently, the task of science is no longer to explain an objective reality, but rather to reduce observations to order finding connections between them and predicting the outcome of new measurements, being aware that single individual events are subject to casuality and only a statistical prediction can be made.

This kind of interpretation first emerged with Bohr's contributions at the International Physics Congress held in Como in September 1927 and  in the next October at the Fifth Solvay Conference in Bruxelles. It  is at the heart of the so-called Copenhagen interpretation developed by Bohr with the contribution of the physicists who visited him in Copenhagen, in particular Heisenberg and Pauli.

According to Max Jammer~\cite{jammer} the writings of Charles Renouvier (1815--1903), \'Emile Boutroux (1845--1921), S\o ren Kierkegaard (1813--1855), and in particular of Harald H\o ffding (1843--1931), with their giving value to the role of consciousness to appreciate the different levels of reality, seem to have been influential in shaping Bohr's philosophical background. In fact,  ``Bohr was primarily a philosopher, not a physicist, but he understood that natural philosophy in our day and age carries weight only if its every detail can be subjected to the inexorable test of experiment''~\cite{rozental}. And with the wave-particle duality of atomic phenomena experiment was unavoidably showing complementary aspects of reality. Certainly, the development of the complementarity idea was tuned with the {\it Zeitgeist\/} of that time, where in any field of the culture between the end of the 18th and the beginning of the 19th century a subjective perspective replaced the objective positivistic one, and the human expression in the arts, literature, philosophy, and ultimately in science became more abstract~\footnote{%
For the interested Italian speaking reader a presentation of the cultural environment in Europe by the time of the rise of quantum mechanics can be found in~\cite{boffi}, where Bohr's and Born's communications at the Como Congress are translated and discussed. Ref.~\cite{boffi} is part of a series of booklets, available at www.pv.infn.it/\lower 3pt\hbox{$\tilde{\ }$}boffi/quaderni.html, where the original papers of those who fabricated quantum mechanics are presented and discussed.
}.

``We might call modern quantum theory as `The Theory of Complementarity' (in analogy with the terminology `Theory of Relativity')''~\cite{handbook}. Gradually convinced that ``it must never be forgotten that we ourselves are both actors and spectators in the drama of existence''~\cite{complem}, Bohr applied the complementarity concept not only to physics, but even in the search for a harmonious attitude towards life.

The idea that the end of the story was achieved with complementarity and completeness of quantum theory was not entirely convincing. Most scientists, also today, do not care about the philosophy of quantum mechanics. Pragmatically, they accept the statistical interpretation of the formalism and apply it  to make predictions. ``Questions of this type appear to be the only ones to which the quantum theory can give a definite answer, and they are probably the only ones to which the physicist requires an answer''~\cite{dirac-interpret}. Therefore, ``ordinary quantum mechanics (as far as I know) is just fine for all practical purposes''~\cite{fapp}. Others, however, like Einstein and Schr\"odinger, tried to envisage paradoxical situations~\cite{epr,cat} to show that quantum theory is not complete and requires further study to understand the measurement process and the consequences of quantum entanglement produced by the linear superposition principle. After all, also the measurement device should be described within quantum theory because it is built by matter ultimately obeying quantum physics. In fact, (local) hidden variables suggested by a causal interpretation of quantum mechanics have to be excluded~\cite{bell,aspect}, while a large debate on measurement theory is still going on involving the decoherence phenomenon which could reconcile quantum and classical behaviour (see, e.g.,~\cite{gimo,decoherence}). However, this is another story which is outside the scope of the present review limited to the rise of quantum mechanics.




\begin{thebibliography}{99}
					
\bibitem{mehra}
\name{J.~Mehra and H.~Rechenberg} \book{Historical Development of Quantum Theory (Springer-Verlag, New York-Heidelberg-Berlin, 1982--1987) volumes 1-5.}

\bibitem{millikan}
\name{R.A.~Millikan} \rev{Phys. Rev.} {7} {355--388} {1916}

\bibitem{lewis}
\name{G.N.~Lewis} \rev{Nature} {118} {874--875} {1926}

\bibitem{BKS}
\name{N.~Bohr, H.A.~Kramers and J.C.~Slater} \rev{Phil. Mag} {47} {785--822} {1924}; \rev{Zeits. f. Phys.} {24} {69--87} {1924}

\bibitem{GB}
\name{W.~Bothe and H.~Geiger} \rev{Naturwiss,} {13} {440--441} {1925}; \rev{Zeits. f. Phys.} {32} {639--663} {1925}

\bibitem{compton}
\name{A.H.~Compton and A.W.~Simon} \rev{Phys. Rev.} {26} {289--299} {1925}

\bibitem{QM}
\name{M.~Born} \rev{Zeits. f. Phys.}{26}{379--395}{1924}

\bibitem{grundpostulate}
\name{N.~Bohr} \rev{Zeits. f. Phys.}{13}{117--165}{1923}

\bibitem{umdeutung}
\name{W.~Heisenberg} \rev{Zeits. f. Phys.} {33}{879--893}{1925}  

\bibitem{drei}
\name{M.~Born and P.~Jordan} \rev{Zeits. f. Phys.}{34} {858--888}{1925};
\name{M.~Born, W.~Heisenberg and P.~Jordan} \rev{Zeits. f. Phys.}{35}{557--615}{1926}

\bibitem{hydrogen}
\name{W.~Pauli} \rev{Zeits. f. Phys.} {36}  {336--363}{1926}

\bibitem{erwin}
\name{E.~Schr\"odinger} \rev{Ann. der Phys.} {79}{361--376} {1926}; \rev{ib.} {79}{489--527}{1926}; \rev{ib.}{80} {437--490}{1926}; \rev{ib.} {81} {109--139}{1926}

\bibitem{kragh}
\name{H.~Kragh} \rev{Am. J.  Phys.} {52} {1024--1033} {1984}

\bibitem{vicki}
\name{W.~Pauli and V.~Weisskopf} \rev{Helv. Phys. Acta} {7}  {709--729} {1934}

\bibitem{wiener}
\name{M.~Born and N.~Wiener} \rev{J. Math. \& Phys. M.I.T.} {5} {84--98} {1926}; \rev{Zeits. f. Phys.} {36} {174--187}{1926}

\bibitem{equivale}
\name{E.~Schr\"odinger} \rev{Ann. der Phys.}  {79} {734--756} {1926} 

\bibitem{waerden}
\name{B.L.~van der Waerden} in \name{J.~Mehra ed.} \book{The Physicist's Conception of Nature (D.~Reidel Publ. Co., Dordrecht-Boston, 1973) p. 276--293.}

\bibitem{lanczos}
\name{C.~Lanczos} \rev{Zeits. f. Phys.} {35} {812--830}{1926}

\bibitem{eckart}
\name{C.H.~Eckart} \rev{Phys. Rev.} {28} {711--726}{1926}

\bibitem{fundamental}
\name{P.A.M.~Dirac} \rev{Proc. Roy. Soc. (London)} {A 109} {642--653}{1925}

\bibitem{prelim}
\name{P.A.M.~Dirac} \rev{Proc. Roy. Soc. (London)} {A 110} {561--579}{1926}

\bibitem{dirac-interpret}
\name{P.A.M.~Dirac} \rev{Proc. Roy. Soc. (London)} {A 113} {621--641}{1927}

\bibitem{dirac-compton}
\name{P.A.M.~Dirac} \rev{Proc. Roy. Soc. (London)} {A 111} {405--423}{1926}

\bibitem{letter-a}
\name{A.H.~Compton} to Dirac, August 26, 1926, in {\it Archive for History of Quantum Mechanics (Niels Bohr Institute, Copenhagen).}

\bibitem{braket}
\name{P.A.M.~Dirac} \rev{Proc. Roy. Soc. (Edinburgh)} {59} {122--129}{1939}

\bibitem{diracbook}
\name{P.A.M.~Dirac}  \book{The Principles of Quantum Mechanics (Clarendon Press, Oxford, 1930)}.

\bibitem{diracstat}
\name{P.A.M.~Dirac} \rev{Proc. Roy. Soc. (London)} {A 112} {661--677} {1926}

\bibitem{neumann}
\name{J.~von Neumann} \book{Mathematische Grundlagen der Quantenmechanik (J.~Springer, Berlin, 1932)}; \book{Mathematical Foundations of Quantum Mechanics (Princeton University Press, Princeton, N.J., 1955)}.

\bibitem{stetige}
\name{E.~Schr\"odinger} \rev{Naturwiss.} {14} {664--666} {1926}

\bibitem{glauber}
\name{R.J.~Glauber} \rev{Phys. Rev.} {131} {2766--2788} {1963}

\bibitem{madelung}
\name{E.~Madelung} \rev{Zeits. f. Phys.} {40} {322--326} {1927}

\bibitem{broglie}
\name{L.~de Broglie} \rev{J. de Phys.} {8} {225--241} {1927}

\bibitem{bohm}
\name{D.~Bohm} \rev{Phys. Rev.} {85} {166--179, 180--193} {1952}

\bibitem{born}
\name{M.~Born} \rev{Zeits. f. Phys.} {40} {167--192} {1927}

\bibitem{stoss}
\name{M.~Born} \rev{Zeits. f. Phys.} {36} {863--867} {1926}; \rev{ib.} {38} {803--827} {1926}

\bibitem{jordan-a}
\name{P.~Jordan} \rev{Zeits. f. Phys.} {37} {383--386} {1926} 

\bibitem{jordan-b}
\name{P.~Jordan} \rev{Zeits. f. Phys.} {40} {809--838} {1927}

\bibitem{london}
\name{F.W.~ London} \rev{Zeits. f. Phys.} {37} {915--925} {1926} 

\bibitem{stoner}
\name{E.C.~Stoner} \rev{Phil. Mag.} {48} {719--726} {1924}

\bibitem{pre-spin}
\name{W.~Pauli} \rev{Zeits. f. Phys.} {31} {765--783} {1925}

\bibitem{spinhyp}
\name{G.E.~Uhlenbeck and S.A.~Goudsmit} \rev{Naturwiss.}{13} {953--954}{1925} 

\bibitem{anomalous}
\name{W.~Heisenberg and P.~Jordan} \rev{Zeits. f. Phys.} {37} {263--277} {1926}

\bibitem{thomas}
\name{L.H.~Thomas} \rev{Nature} {117} {514} {1926}

\bibitem{lamb}
\name{W.E.~Lamb, jr. and R.C.~Retherford} \rev{Phys. Rev.} {72} {241--243} { 1947}; 
\rev{ib.} {79} {549--572} {1950}; \rev{ib.} {81} {222--232} {1951}

\bibitem{spin}
\name{W.~Pauli} \rev{Zeits. f. Phys.} {43}{601--623}{1927}

\bibitem{electron}
\name{P.A.M.~Dirac} \rev{Proc. Roy. Soc. (London)} {A 117} {610--624} {1928}; \rev{ib.} {A 118} {351--361}{1928} 

\bibitem{bose}
\name{S.N.~Bose} \rev{Zeits. f. Phys.} {26} {178--181} {1924}

\bibitem{einsteinstat}
\name{A.~Einstein} \rev{Sitzungsber. Preuss. Akad. Wiss. (Berlin)} {22} {261--267} {1924};  \rev{ib.}{23} {3--14} {1925};  \rev{ib.}{23} {18--25} {1925}

\bibitem{helium}
\name{W.~Heisenberg} \rev{Zeits. f. Phys.} {39} {499--518} {1926}

\bibitem{hund}
\name{F.~Hund} \rev{Zeits. f. Phys.} {33} {347--371} {1925}

\bibitem{fermi}
\name{E.~Fermi} \rev{Rend. Acc. Lincei} {3} {145--149}{1926}; \rev{Zeits. f. Phys.} {36} {902--912} {1926}

\bibitem{slater}
\name{J.C.~Slater} \rev{Phys. Rev.} { 34} {1293--1322} {1929}

\bibitem{fock}
\name{V.A.~Fock} \rev{Zeits. f. Phys.} {61} {126--148} {1930}

\bibitem{hartree}
\name{D.R.~Hartree} \rev{Proc. Cambridge Phil. Soc.} {24} {89--110, 111--132} {1928}

\bibitem{condonshortley}
\name{E.U.~Condon and G.H.~Shortley} \book{Theory of Atomic Spectra (Cambridge University Press, Cambridge, 1935).}

\bibitem{pauli}
\name{W.~Pauli} \rev{Zeits. f. Phys.}{41} {81--102}{1927}

\bibitem{pauli-stat}
\name{W.~Pauli} \rev{Phys. Rev.} {58} {716--722} {1940}

\bibitem{indetermina}
\name{W.~Heisenberg} \rev{Zeits. f. Phys.} {43} {172--198}{1927} 

\bibitem{letter-b}
\name{W.~Heisenberg} to Pauli, June 8, 1926, in \name{A.~Hermann, K. von Meyenn, V.F.~Weisskopf, eds.} \book{Wolfgang Pauli: Wissenschaftlicher Briefwechsel mit Bohr, Einstein, Heisenberg, u.a. , vol. 1: 1919-1929 (Springer, Berlin, 1979, p. 328--329).}

\bibitem{kennard}
\name{E.H.~Kennard} \rev{Zeits. f. Phys.} {44} {326--352} {1927}

\bibitem{condon}
\name{E.U.~Condon} \rev{Science} {69} {573-574} {1929}

\bibitem{robertson}
\name{H.P.~Robertson} \rev{Phys. Rev.} {34} {163--164} {1929}

\bibitem{prussian}
\name{E.~Schr\"odinger} \rev{Sitzungsber. Preuss. Akad. Wiss. (Berlin)} {24} {296--303} {1930}

\bibitem{stapp}
\name{H.P.~Stapp} \rev{Am. J. Phys.} {40} {1098--1116} {1972}

\bibitem{order}
\name{N.~Bohr} \book{Atomic Theory and the Description of Nature (Cambridge University Press, Cambridge, 1934, p. 3).}

\bibitem{fenomeno}
\name{N.~Bohr} \book{Essays 1958--1962 in Atomic Physics and Human Knowledge (Wiley, New York, 1963, p. 3).}

\bibitem{jammer}
\name{M.~Jammer} \book{The conceptual development of quantum mechanics (McGraw-Hill, New York, 1966; 2nd ed., American Institute of Physics, New York, 1989)}.

\bibitem{rozental}
\name{W.~Heisenberg} in \name{S.~Rozental, ed.} \book{Niels Bohr: His life and work as seen by his friends and colleagues (North-Holland, Amsterdam, 1967, p. 94--108).}

\bibitem{boffi}
\name{S.~Boffi} \book{Il postulato dei quanti e il si\-gnificato della funzione d'onda (Bibliopolis, Napoli, 1996)}.

\bibitem{handbook}
\name{W.~Pauli} \book{Handbuch der Physik, Vol. 5, Part I: Prinzipien der Quantentheorie I (Springer-Verlag, Berlin, 1958).}

\bibitem{complem}
\name{N.~Bohr} \rev{Naturwiss.} {18} {73--78} {1930}

\bibitem{fapp}
\name{J.S.~Bell} in \name{A.I.~Miller, ed.} \book{Sixty-Two Years of Uncertainty. Historical, Philosophical, and Physical Inquiries into the Foundations of Quantum Mechanics (Plenum Press, New York, 1989, p. 17--31).}

\bibitem{epr}
\name{A.~Einstein, B.~Podolsky and N.~Rosen} \rev{Phys. Rev.} {47} {777--780} {1935}

\bibitem{cat}
\name{E.~Schr\"odinger}  \rev{Naturwiss.} {23} {807--812, 823--828, 844--849} {1935}

\bibitem{bell}
\name{J.S.~Bell} \rev{Physics} {1} {195--200} {1964}

\bibitem{aspect}
\name{A.~Aspect, J.~Dalibard and G.~Roger} \rev{Phys. Rev.} {47} {460--463} {1981}; \rev{ib.} {49} {91--94, 1804--1807} {1982}

\bibitem{gimo}
\name{E.B.~Karlsson and E.~Br\"andas, eds.} \book{Modern Studies of Basic Quantum Concepts and Phenomena, Proceedings of Nobel Symposium 104}, in \rev{Physica Scripta} {T76} {1--232} {1998}

\bibitem{decoherence}
\name{E.~Joos, H.D.~Zeh, C.~Kiefer, D.~Giulini, J,~Kupsch and I.-O.~Stamatescu eds.} \book{Decoherence and the Appearance of a Classical World in Quantum Theory (Springer, Berlin, 2003)}

\end{thebibliography}
\end{document}